\documentclass[twocolumn,aps,preprintnumbers,amsmath,amssymb,superscriptaddress,nobibnotes]{revtex4}
\usepackage{graphicx}
\usepackage{dcolumn}
\usepackage{bm}
\usepackage{graphics}
\usepackage{longtable}
\usepackage{units}
\usepackage{textcomp}
\usepackage{float}

\begin{document}

\title{Multi-frequency perturbations in matter-wave interferometry}

\author{A. G\"{u}nther}
\email{a.guenther@uni-tuebingen.de}
\affiliation{Institute of Physics and Center for Collective Quantum Phenomena in LISA$^+$,
University of T\"{u}bingen, Auf der Morgenstelle 14, 72076 T\"{u}bingen, Germany}
\author{A. Rembold}
\affiliation{Institute of Physics and Center for Collective Quantum Phenomena in LISA$^+$,
University of T\"{u}bingen, Auf der Morgenstelle 15, 72076 T\"{u}bingen, Germany}
\author{G. Sch\"{u}tz}
\affiliation{Institute of Physics and Center for Collective Quantum Phenomena in LISA$^+$,
University of T\"{u}bingen, Auf der Morgenstelle 15, 72076 T\"{u}bingen, Germany}
\author{A. Stibor}
\email{alexander.stibor@uni-tuebingen.de}
\affiliation{Institute of Physics and Center for Collective Quantum Phenomena in LISA$^+$,
University of T\"{u}bingen, Auf der Morgenstelle 15, 72076 T\"{u}bingen, Germany}

\begin{abstract}
High contrast matter-wave interferometry is essential in various fundamental quantum mechanical experiments as well as for technical applications. Thereby, contrast and sensitivity are typically reduced by decoherence and dephasing effects. While decoherence accounts for a general loss of quantum information in a system due to entanglement with the environment, dephasing is due to collective time-dependent external phase shifts, which can be related to temperature drifts, mechanical vibrations or electromagnetic oscillations. In contrast to decoherence, dephasing can in principle be reversed. Here, we demonstrate in experiment and theory a method for the analysis and reduction of the influence of dephasing noise and perturbations consisting of several external frequencies in an electron interferometer. This technique uses the high spatial and temporal resolution of a delay line detector to reveal and remove dephasing perturbations by second order correlation analysis. It allows matter-wave experiments under perturbingly lab conditions and can be applied in principle to electron, atom, ion, neutron and molecule interferometers.
\end{abstract}

\maketitle

\section{Introduction}
Matter-wave interferometers accomplish an outstanding sensitivity for high precision measurements. This has been shown in various fields, e.g.~by the measurement of inertial forces \cite{gustavson1997,Hasselbach1993}, the gravitational acceleration \cite{Peters1999} and atomic or molecular polarizabilities \cite{Ekstrom1995,Berninger2007}. As a consequence of the high sensitivity, interferometers are also susceptible to external perturbations and noise that dephase the matter-wave. Typical dephasing sources are temperature drifts, mechanical vibrations or electromagnetic oscillations. These influences result in a ''washing-out'' of the interference fringes. It generally reduces the interference contrast \cite{Stibor2005} and is especially an obstacle in precision phase measurements such as in Aharonov-Bohm physics \cite{Aharonov1959,Schuetz2015b}. Dephasing also needs to be prevented in decoherence studies  where the transition from quantum to classical behaviour is analyzed through the contrast loss in matter-wave interferometers \cite{Hornberger2003,Hackermuller2004,Sonnentag2007}. Decoherence destroys the phase relation between two beam paths by interaction with the environment, whereas dephasing is due to a collective time-dependent phase shift of the wave functions. Both reduce the contrast and need to be distinguished to study experimentally the theory of decoherence \cite{Zurek2003}. The knowledge of decoherence mechanisms is of importance to achieve long coherence times in a variety of hybrid quantum devices in the community to date \cite{Wallquist2009,Sague2007,Andre2006,Hunger2010,Chan2011,Gierling2011}.

Certainly, isolating, damping or shielding of interferometers reduce the dephasing influences. However, in a perturbingly environment this often cannot be achieved efficiently in a broad frequency range. In addition, for some applications such as mobile interferometric devices \cite{Hauth2013}, the perturbation sources and therefore the shielding requirements are changing between different experimental runs.

In this article we present a general method to reveal dephasing and reduce the influence of external multi-frequency perturbations and noise by second order correlation analysis of an interference pattern. The technique can in principle be applied in all matter-wave interferometers, where the detector has a high spatial and temporal resolution. Such detectors are available for electrons \cite{Jagutzki2002}, ions \cite{Jagutzki2002}, neutrons \cite{Siegmund2007} and neutral atoms \cite{Schellekens2005}. Perturbation frequencies below the mean particle count rate can then be revealed after data acquisition is completed.

Recently, we demonstrated \cite{Rembold2014} that the influence of a single perturbation frequency can be isolated, removed and the complete spatial interference pattern be recovered. All parameters such as the contrast, pattern periodicity, perturbation frequency and amplitude can be extracted from the perturbed interference data. This is useful to eliminate a distinct electromagnetic oscillation such as \unit[50]{Hz} noise from the electric network or a vibration from a rotating vacuum pump.

In this article we enhance this method towards several frequencies with different amplitudes, that dephase the interference pattern at once. We provide a detailed theoretical description of the multi-frequency case, for which we calculate an exact solution for the second order correlation function. While the theory holds for an arbitrary number of discrete perturbation frequencies, it requires increasing computational power. However, an approximate solution can be found, which is used to extend the theory to arbitrary noise spectra with a high-frequency cutoff. Such noise is often found in typical lab situations, where several low frequency noise sources disturb the interference pattern at the same time. Here, the relative phases of the frequency components do not influence the correlation function and the approximate solution is applicable.

We demonstrate our method by the application of an electron matter-wave interferometer in combination with a delay-line detector. Using an external magnetic field, we introduce artificial multi-frequency noise, which we analyze on the basis of the "washed-out" interference pattern. In detail, we demonstrate two- and three-frequency perturbations, including the special case, where the frequencies are multiples of each other. The influence of broad-band noise is investigated by using a perturbation spectrum with a cutoff frequency of $\unit[150]{Hz}$. For all scenarios, the measured particle correlations are compared to our theory.

\section{Theory}
Recently, we demonstrated a full correlation analysis for a single frequency perturbation of a matter-wave interference pattern in an interferometer \cite{Rembold2014}. In the following we discuss the general case and extend the correlation analysis to multiple perturbation frequencies and broad-band noise.

Similar as before, we start with the probability distribution $f(y,t)$ of particle impacts at the detector
\begin{equation}
f(y,t) = f_0\left(1+K\cos \left(ky + \varphi\left(t\right)\right)\right) \label{eq1} ~.
\end{equation}
Here $K$ denotes the contrast, $k=2\pi/\lambda$ the wave number of the unperturbed interference pattern with spatial periodicity $\lambda$, and $f_0$ assures normalization. The perturbation is mediated via the time dependent phase shift $\varphi(t)$, which we assume as superposition of $N$ discrete harmonic frequencies
\begin{equation}
\varphi(t)=\sum_{j=1}^N \varphi_j\cos\left(\omega_j t + \phi_j\right) ~. \label{eq2}
\end{equation}
Thereby $\varphi_j$ and $\phi_j$ denote the peak phase deviation and the perturbation phase at the frequency component $\omega_j$. These perturbations will typically lead to a "wash-out" of the time-averaged interference pattern at the detector.

To regain information about the unperturbed interference pattern, the spatial and temporal information of the particles arriving at the detector is recorded and evaluated by correlation analysis. Using Eq. (\ref{eq1}), the second-order correlation function reads
\begin{equation}
g^{(2)}(u,\tau) = \frac{\ll f(y+u,t+\tau) f(y,t)\gg_{y,t}}{\ll f(y+u,t+\tau) \gg_{y,t} \ll f(y,t) \gg_{y,t}}~,
\label{eq3}
\end{equation}
with $\ll \cdot \gg_{y,t}$ denoting the average over position and time
\begin{equation}
\ll f(y,t) \gg_{y,t} = \lim_{Y,T\rightarrow\infty} \frac{1}{TY}\int_0^T \int_{-Y/2}^{Y/2} f(y,t) \,dydt ~. \label{eq4}
\end{equation}
In the limit of large acquisition times $T\gg 2\pi/\omega_j$ and lengths $Y\gg 1/k$, Eq. (\ref{eq3}) can be solved analytically yielding
\begin{equation}
g^{(2)}(u,\tau) = 1 + \frac{1}{4}K^2\left[\mbox{e}^{-iku}A_{+} + \mbox{e}^{iku}A_{-}\right] ~,\label{eq5}
\end{equation}
with
\begin{eqnarray}
A_{\pm} &=& \prod_{j=1}^N\sum_{n_j,m_j}\!J_{n_j}\left(\varphi_j\right)\!J_{m_j}\left(\varphi_j\right)\mbox{e}^{i m_j\omega_j\tau}\chi_{n_j,m_j}^\pm\label{eq6}\\
\chi_{n_j,m_j}^\pm &=& \mbox{e}^{i\left[n_j\left(\phi_j\pm\frac{\pi}{2}\right)+m_j\left(\phi_j\mp\frac{\pi}{2}\right)\right]} ~, \label{eq7}
\end{eqnarray}
where $J_n$ denote the Bessel functions of first kind and the sum in Eq. (\ref{eq6}) has to be taken over all integer-duplets $\left\{n_j, m_j\right\}\in\mathbb{Z}, j=1\ldots N$, for which
\begin{equation}
\sum_j\left(n_j+m_j\right)\omega_j=0 \label{eq8}
\end{equation}
is fulfilled. Although this makes in principle for an infinite number of addends, the contribution of integer-duplets $\left\{n_j,m_j\right\}>\phi_j$ to $A_{\pm}$ is suppressed due to the strong decay of the Bessel functions.

For multiple perturbation frequencies $\omega_j$ and broad-band noise spectra, the constraint from Eq. (\ref{eq8}) requires usually $m_j=-n_j$ for all $j=1\ldots N$. With $\chi_{n_j,-n_j}^\pm=(-1)^{n_j}$, this results in a correlation function
\begin{equation}
g^{(2)}(u,\tau) = 1 + A\left(\tau\right)\cos\left(ku\right) ~,\label{eq9}
\end{equation}
with
\begin{equation}
A(\tau) = \frac{1}{2}K^2\prod_{j=1}^N \underbrace{\sum_{n_j=-\infty}^\infty J_{n_j}\left(\varphi_j\right)^2 \mbox{e}^{-in_j\omega_0\tau}}_{J_0\left(\varphi_j\right)^2 + 2\sum_{n_j=1}^\infty J_{n_j}\left(\varphi_j\right)^2\cos\left(n_j\omega_j\tau\right)} ~.\label{eq10}
\end{equation}

The correlation function for multiple dephasing oscillations will thus show a periodic modulation in the spatial distance $u$ between two detection events. It has the same periodicity as the undisturbed interference pattern. Therefore, it can proof matter-wave interference in a situation where the periodic pattern would be "washed-out" after signal integration. The amplitude $A(\tau)$ of this modulation, however, depends on the correlation time $\tau$, and is given by the specific perturbation spectrum. The function $A(\tau)$ includes a superposition of sidebands at discrete frequencies $n_j\omega_j$ ($n_j\in \mathbb{Z}$). Their strengths are defined by the peak phase deviation $\varphi_j$ via the Bessel functions $J_{n_j}(\varphi_j)$. Maximal modulation amplitude of $0.5\,K^2$ is achieved at correlation time $\tau=0$.

While for $N=1$, the results from Eq. (\ref{eq9}) and (\ref{eq10}), coincide with our previous result found for single frequency perturbations \cite{Rembold2014}, special care has to be taken for small numbers of perturbation frequencies, which are multiples of each others. In this case the constraint from Eq. (\ref{eq8}) might be fulfilled for integers $m_j\neq-n_j$, leading to additional terms in Eq. (\ref{eq10}). This can be easily seen in the two-frequency case, with $\omega_2=2\omega_1$, where Eq. (\ref{eq8}) is not only solved for $m_{1/2}=-n_{1/2}$, but for all multiplets with $n_1+m_1=-2\left(n_2+m_2\right)$, such as  $\left\{n_1,m_1,n_2,m_2\right\}=\left\{2,2,-1,-1\right\}$. Following Eq. (\ref{eq5})--(\ref{eq7}), additional terms appear in the correlation function, with $\chi_{n_j,m_j}^\pm\neq\pm1$, thus introducing a phase shift to the correlation pattern, which depends on the correlation time and the perturbation phases $\phi_j$. The correlation function can then not be described by the simple form of Eq. (\ref{eq9}), but has to be calculated via Eq. (\ref{eq5})--(\ref{eq7}).

Similar effects may arise for larger numbers of perturbation frequencies, but the contribution of these parts become more and more negligible with respect to the main correlation component described by Eq. (\ref{eq9})--(\ref{eq10}).

Examples for all of the above scenarios will be given in section \ref{sec4}.

\section{Experiment}
To demonstrate our method we use an electron biprism interferometer and analyze interference patterns that were subject to multi-frequency perturbations. The experimental setup is equal to the one used for the single frequency correlation analysis and is described elsewhere \cite{Rembold2014,Schuetz2014,Hasselbach1998a,Maier1997,Hasselbach2010}. A coherent beam of electrons is field emitted from a single atom tip cathode \cite{Kuo2006a,Kuo2008} and guided towards an electrostatic biprism wire. Being on a positive potential it separates and recombines the matter-wave leading to an interference pattern \cite{Mollenstedt1956a}. After magnification the interferogram is amplified by two multichannel plates and detected by a delay line detector with high spatial and temporal resolution \cite{Jagutzki2002}. The whole setup is in an UHV-chamber at a pressure of \unit[$1.4\times 10^{-10}$]{mbar}. The beam path inside the chamber is shielded by a mu-metal tube.

For the course of this paper, we artificially disturb the pattern by several superposed oscillating magnetic fields. They are generated by two external magnetic coils in Helmholtz configuration positioned around the vacuum chamber. The frequencies are introduced by a frequency generator. The oscillating magnetic field shifts the interference pattern perpendicular to the fringes and the integrated pattern gets "washed-out".

The extraction of the second order correlation function from the detector signal data was performed according to the same procedure as described in \cite{Rembold2014}. From the detector we get the position $y_i$ and the time $t_i$ of all particle events $i=1...M$. We then determine the number $M_{u,\tau}$ of particle pairs $(i,j)$ with $(y_i-y_j)\in[u,u+\Delta u]$ and $(t_i-t_j)\in[\tau,\tau+\Delta \tau]$, yielding the correlation function
\begin{equation}
g^{(2)}(u,\tau) = \frac{TY}{M^2\Delta\tau\Delta u}\frac{M_{u,\tau}}{\Big(1-\frac{\tau}{T}\Big)\Big(1-\frac{|u|}{Y}\Big)} ~.
\label{eq11}
\end{equation}
Here, $\Delta \tau$ and $\Delta u$ denote the discretization step size in correlation time and position, respectively. $TY/M^2$ is a normalization factor and $[(1-\tau/T)(1-|u|/Y)]^{-1}$ corrects $M_{u,\tau}$ for the finite acquisition time $T$ and length $Y$.

\begin{figure}
\includegraphics[width=0.45\textwidth]{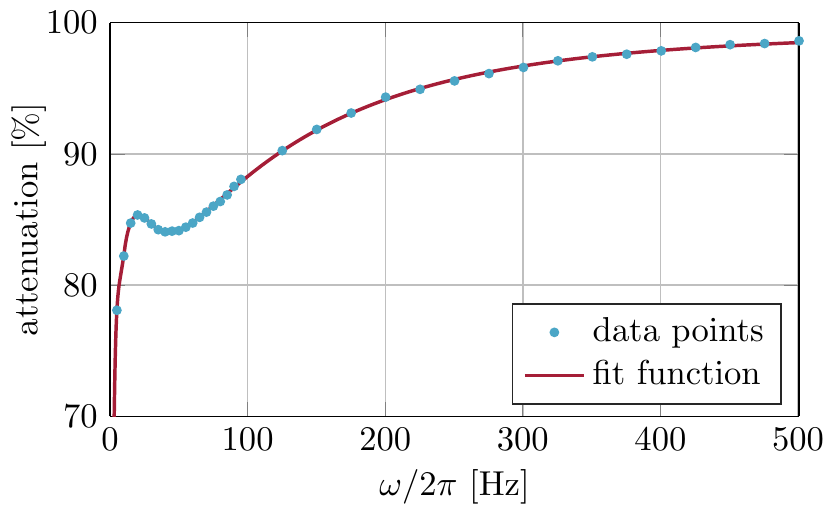} \caption{(Color online) Frequency dependent attenuation of the mu-metal shield around the beam path. The data was obtained by applying single frequency perturbations between $0$ and $\unit[500]{Hz}$ to the interferogram. Each point corresponds to a correlation analysis of a disturbed interference pattern where the extracted perturbation amplitude has been set in relation to the phase shift at constant magnetic field ($\unit[0]{Hz}$). The attenuation curve has been fitted by an appropriate function.}
\label{fig1}
\end{figure}

Before applying multi-frequency noise, the mu-metal shielding has to be characterized, as it causes a frequency dependent damping of the externally applied fields. This is done by applying a single frequency perturbation via the external coils with the perturbation frequency stepwise increased from $0$ to $\unit[500]{Hz}$. For each measurement the same oscillation amplitude was applied to the coils and an electron interferogram was recorded. The peak phase deviation of the perturbation in the individual interference pattern was extracted by fitting the correlation theory from Eq. (\ref{eq9})--(\ref{eq10}), for the single frequency case ($N=1$), to the measured particle correlations. The resulting phase deviations $\varphi_1(\omega)$ were compared to the phase shift for constant magnetic field $\varphi_1(0)$, which can be extracted from the $\unit[0]{Hz}$ measurement and is uninfluenced from the mu-metal shield. The frequency dependent attenuation of the mu-metal shield is then given by $1-\varphi_1(\omega)/\varphi_1(0)$ and shown in Fig. \ref{fig1} together with an appropriate fit function.

\section{Results}\label{sec4}
For the experimental verification of our multi-frequency correlation theory we investigate different perturbation scenarios. Starting with two and three perturbation frequencies, we also analyze a full perturbation noise spectrum. For all measurements an interference pattern with $7\times 10^5$ counts was recorded.

\begin{figure}
\includegraphics[width=0.475\textwidth]{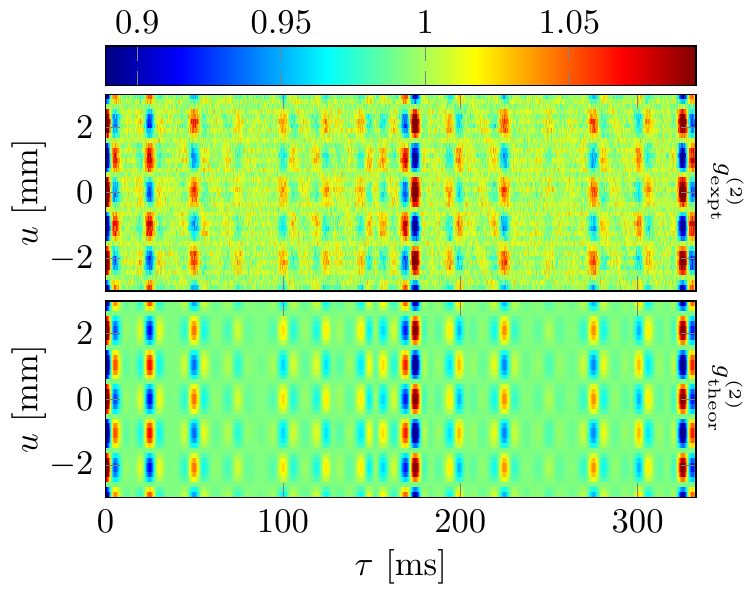} \caption{(Color online) top: Experimental correlation function $g_{\text{expt}}^{(2)}(u,\tau)$ with two perturbation frequencies of $\omega_1 = \unit[2\pi\times 6]{Hz}$ and $\omega_2 = \unit[2\pi\times 40]{Hz}$. bottom: Theoretical fit according to Eq. (\ref{eq5})--(\ref{eq7}) with $N=2$.}
\label{fig2}
\end{figure}

In the first measurement the interference pattern was perturbed by two superposed magnetic oscillations at $\omega_1 = \unit[2\pi\times 6]{Hz}$ and $\omega_2 = \unit[2\pi\times 40]{Hz}$. Using eq. 11, we then extracted the second order correlation function $g_{\text{expt}}^{(2)}(u,\tau)$ from these data, which is shown in the upper panel of Fig.~\ref{fig2}. The theoretical correlation function $g_{\text{theor}}^{(2)}(u,\tau)$, in the lower panel of Fig.~\ref{fig2}, was fitted to the experimental data according to Eq. (\ref{eq5})--(\ref{eq7}) with $N=2$ and shows good agreement with the experiment. From the theoretical fit we extracted the contrast $ K = \unit[62.9]{\%}$, the pattern periodicity $\lambda = \unit[2.08]{mm}$ and the peak phase deviations $\varphi_1 = \unit[1.34]{\pi}$ and $\varphi_2 = \unit[0.93]{\pi}$. The phases $\phi_1$ and $\phi_2$ have no influence at the resulting theoretical function in Fig.~\ref{fig2}, because the constraint from Eq. (\ref{eq8}) is only fulfilled for $m_j=-n_j$ and therefore Eq. (\ref{eq7}) becomes $\chi_{n_j,-n_j}^\pm=(-1)^{n_j}$. In this case we also could have used Eq. (\ref{eq9})--(\ref{eq10}) to fit the experimental data, yielding the same result.

\begin{figure}
\includegraphics[width=0.475\textwidth]{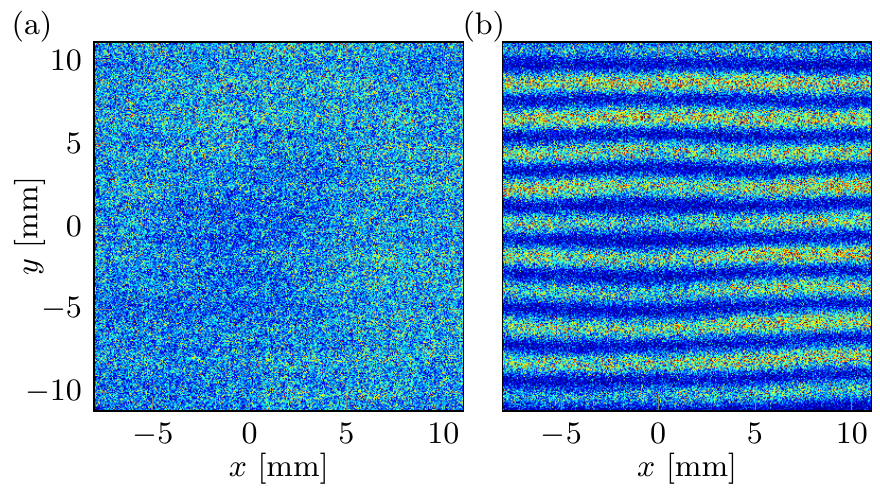} \caption{(Color online) (a) Electron interference pattern perturbed by two frequencies $\omega_1 = \unit[2\pi\times 6]{Hz}$ and $\omega_2 = \unit[2\pi\times 40]{Hz}$. The remaining contrast is $ K = \unit[6.4]{\%}$. (b) Recalculated interference pattern where the dephasing could be corrected by the presented correlation analysis together with Eq. (\ref{eq12}) for $N=2$. The contrast of the reconstructed pattern is $ K = \unit[62.4]{\%}$.}
\label{fig3}
\end{figure}

With the obtained parameters from the theoretical fit it is possible to recalculate the interference pattern from the perturbed one in Fig.~\ref{fig3}(a). To get the unperturbed interference pattern we have to calculate the new positions $y_{new}$ according to the extracted perturbation frequencies $\omega_j$ and peak phase deviations $\varphi_j$
\begin{equation}
\begin{split}
y_{new} = y \,-\, \frac{\lambda}{2\pi}\sum_{j=1}^N \varphi_j \cos\left(\omega_j t + \phi_j\right)~, \label{eq12}
\end{split}
\end{equation}
with the particle positions of the perturbed pattern $y$, pattern periodicity $\lambda$ and the phases of the perturbations $\phi_j$. As the phases could not be obtained from the correlation analysis, they have to be varied to find the maximum contrast of the interference pattern. For our measurement we found $\phi_1 = \unit[-0.33]{\pi}$ and $\phi_2 = \unit[-0.63]{\pi}$. The resulting back-calculated interference pattern is shown in Fig.~\ref{fig3}(b). In the dephased pattern only a small contrast of \unit[6.4]{\%} can be adumbrated. After recalculating the new particle positions the interference pattern is clearly visible and the contrast is significantly higher and amounts \unit[62.4]{\%}.

\begin{figure}
\includegraphics[width=0.475\textwidth]{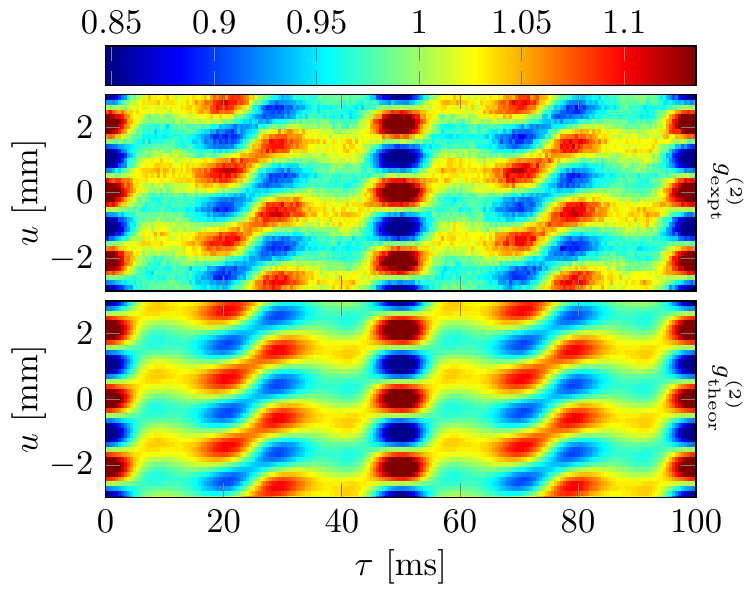} \caption{(Color online) top: Electron interference measurement with two perturbation frequencies, that are multiples of each other, $\omega_1 = \unit[2\pi\times 20]{Hz}$ and $\omega_2 = \unit[2\pi\times 40]{Hz}$. bottom: The exact solution from Eq. (\ref{eq5})--(\ref{eq7}) with $N=2$ was fitted to the experimental correlation data, showing good agreement.}
\label{fig4}
\end{figure}

In a second experiment the correlation function was studied for the case of two perturbation frequencies that are multiples of each other. With $\omega_1 = \unit[2\pi\times 20]{Hz}$ and $\omega_2 = \unit[2\pi\times 40]{Hz}$ the resulting correlation pattern is shown in Fig. \ref{fig4}. As discussed in the theory section, the constraint  from Eq. (\ref{eq8}) is now not only fulfilled for $m_{1/2}=-n_{1/2}$. Thus additional terms appear in the correlation function, that result in a phase shift of the correlation pattern depending on the correlation time $\tau$ and the perturbation phases $\phi_j$. This phase shift is clearly seen in the correlation pattern of Fig. \ref{fig4}. In this case the full theory from Eq. (\ref{eq5})--(\ref{eq7}) is required to fit the experimental data, as the correlation time dependent phase shifts are not included in the approximate solution from eq. (\ref{eq9})--(\ref{eq10}). The fit is shown in the lower panel of Fig. \ref{fig4} and shows good agreement with the experiment. From the fit we extract the contrast $ K = \unit[62.9]{\%}$, the pattern periodicity $\lambda = \unit[2.08]{mm}$, the peak phase deviations $\varphi_1 = \unit[0.50]{\pi}$ and $\varphi_2 = \unit[0.52]{\pi}$ and the phases $\phi_1 = \unit[-0.21]{\pi}$ and $\phi_2 = \unit[0.24]{\pi}$.

\begin{figure}
\includegraphics[width=0.475\textwidth]{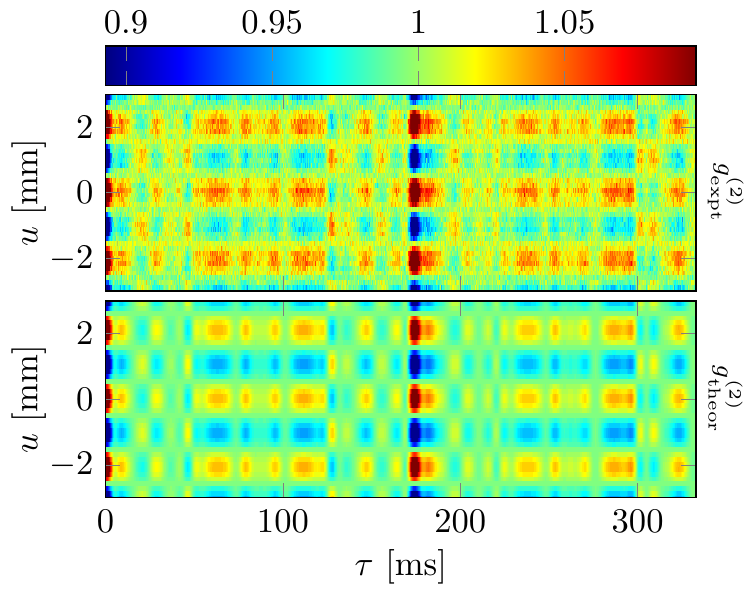} \caption{(Color online) top: Measurement with three perturbation frequencies of $\omega_1 = \unit[2\pi\times 6]{Hz}$, $\omega_2 = \unit[2\pi\times 23]{Hz}$ and $\omega_3 = \unit[2\pi\times 40]{Hz}$. bottom: Fit from Eq. (\ref{eq5})--(\ref{eq7}) with $N=3$ to the experimental correlation data.}
\label{fig5}
\end{figure}

Also the case of three superposed perturbation frequencies was studied and compared to our theoretical approach. The outcome of a corresponding electron interference experiment is shown in Fig.~\ref{fig5} at the top. Here, the frequencies $\omega_1 = \unit[2\pi\times 6]{Hz}$, $\omega_2 = \unit[2\pi\times 23]{Hz}$ and $\omega_3 = \unit[2\pi\times 40]{Hz}$ have been applied. For fitting these data we use the exact solution from Eq. (\ref{eq5})--(\ref{eq7}) for three frequencies, because the constraint from Eq. (\ref{eq8}) is fulfilled for many multiplets $\left\{n_j,m_j\right\}$, with $m_j\neq-n_j$. As for the case of two frequencies, that are multiples of each other (Fig.~\ref{fig4}), this leads to additional terms in the correlation function. The result of the fit is shown in the lower panel of Fig. \ref{fig5}, yielding the contrast $ K = \unit[61.3]{\%}$, the pattern periodicity $\lambda = \unit[2.08]{mm}$, the peak phase deviations $\varphi_1 = \unit[0.76]{\pi}$, $\varphi_2 = \unit[1.01]{\pi}$ and $\varphi_3 = \unit[0.52]{\pi}$ and the phases $\phi_1 = \unit[0.01]{\pi}$, $\phi_2 = \unit[1.01]{\pi}$ and $\phi_3 = \unit[0.01]{\pi}$. With the exact solution a good agreement between theory and experiment can be achieved.

\begin{figure}
\includegraphics[width=0.45\textwidth]{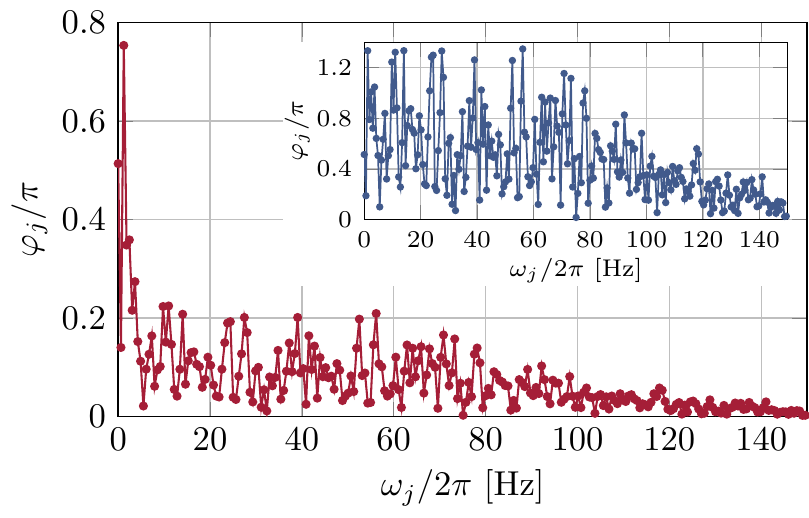} \caption{(Color online) Perturbation spectrum at the position of the electrons which was used to calculate the theoretical correlation function in Fig. \ref{fig6}, according to Eq. (\ref{eq9})--(\ref{eq10}). The spectrum originates from the externally applied noise spectrum (see inset) and the frequency dependent damping of the mu-metal shield (see Fig. \ref{fig1}).}
\label{fig6}
\end{figure}

\begin{figure}
\includegraphics[width=0.475\textwidth]{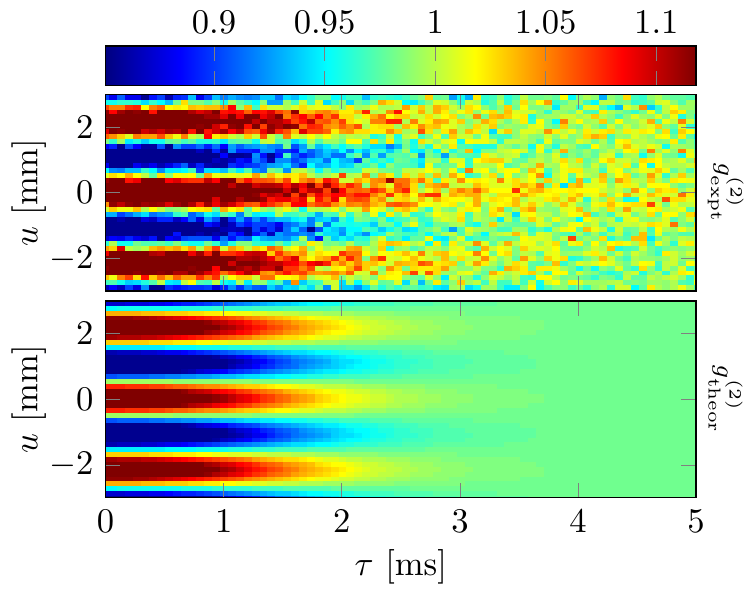} \caption{(Color online) Correlation analysis of an interference pattern that was perturbed by a broad-band noise spectrum with a cutoff frequency at $\unit[150]{Hz}$. This simulates a perturbingly lab environment. top: correlation function $g_{\text{expt}}^{(2)}(u,\tau)$ as obtained from the measurement. bottom: Theoretical fit according to Eq. (\ref{eq9})--(\ref{eq10}).}
\label{fig7}
\end{figure}

In the last experiment we investigate a perturbation consisting of a broad frequency spectrum. This should simulate typical noise from a lab environment originating from electric or magnetic devices or mechanical vibrations. We introduced a perturbation noise spectrum from $\unit[0]{Hz}$ up to a cutoff frequency at $\unit[150]{Hz}$ to the magnetic coils. A part of the time signal of the applied perturbation spectrum was measured while integrating the signal at the detector. The Fourier spectrum of this time signal (see inset of Fig. \ref{fig6}) was attenuated according to the fit function shown in Fig.~\ref{fig1} and the amplitudes of the frequency components at the positions of the particles were calculated. The result is shown in Fig.~\ref{fig6}. These values were used to calculate the theoretical correlation function according to Eq. (\ref{eq9})--(\ref{eq10}). In this case it is possible to use the approximate solution for the correlation function, because the perturbation consists of a large number of frequencies and uncorrelated phases. Fig. \ref{fig7} shows the measured data together with the theoretical fit, from which we deduce the contrast of the interference pattern $K = \unit[63.4]{\%}$ and the pattern periodicity $\lambda = \unit[2.16]{mm}$. Again the experimental and theoretical correlation functions agree well.

\section{Conclusion}
Matter-wave interference experiments appear in various fields of experimental quantum physics for atoms \cite{Andrews1997}, molecules \cite{Brezger2002}, neutrons \cite{Rauch2015}, electrons \cite{Hasselbach2010}, and ions \cite{Hasselbach1998a}. Most of these approaches have a spatial resolution for the interference pattern, but not a temporal one. Therefore, these kind of setups are susceptible to dephasing effects. Oscillations from electromagnetic sources, temperature drifts or mechanical vibrations decrease the contrast or totally "wash-out" the interference fringes. For that reason good shielding and damping is required.

However, the technological progress allows for particle detectors with a high spatial and temporal resolution. In this paper we demonstrated that the additional temporal information can be used to determine and decrease the effect of dephasing significantly even if the interferometer is under the influence of a multi-frequency perturbation and broad-band noise. We presented a theoretical approach that uses second order correlation theory to analyze an interference pattern disturbed by a spectrum of various frequencies with different amplitudes and phases. Thereby, the specific composition of the perturbation frequencies is relevant. It determines, whether the exact solution is needed, or an approximate solution, requiring less computational effort, is sufficient.

Our method was verified experimentally by electron interference patterns that were artificially perturbed and "washed-out" by several superposed frequencies. The resulting correlation data for two and three frequency perturbations and a perturbation noise spectrum matched well with our theoretical description. We demonstrated the reconstruction of the unperturbed interference pattern with the obtained information about the perturbation frequencies and amplitudes.

The method can be applied if the perturbation frequencies are lower than the average count rate of the detected particles. This requirement is fulfilled for most interferometry experiments, where a count rate of several kHz and perturbation frequencies of a few hundred Hz are common. The obtained information from our analysis allows for the optimization of damping and shielding installations. Our technique has a potential application for (mobile) interferometers in a noisy environment and in sensor technology. The susceptibility of electron interferometers may be used for the construction of a sensor with a high sensitivity for electromagnetic and vibrational frequencies. The resolution of such a device would presumably only be limited by the count rate of the particles and the distance of the interference fringes. It can be increased by the use of heavier particles, such as ions, or a larger beam path separation \cite{Clauser1988}.

Our method identifies and analyzes multi-frequency noise in interferometric measurements. It is applicable for a broad variety of matter-wave interferometers, where the reduction of dephasing and the proof of the wave nature of particles is required in a noisy environment with multi-frequency perturbations.\\

\section{Acknowledgements}
This work was supported by the Deutsche Forschungsgemeinschaft (DFG, German Research Foundation) through the Emmy Noether program STI 615/1-1. A.G. acknowledges support from the DFG SFB TRR21, and A.R. from the Evangelisches Studienwerk e.V. Villigst. The authors thank A. Pooch and R. R\"opke for helpful discussions.

\end{document}